\documentclass[aps,prl,floatfix,showpacs,tightenlines,twocolumn,superscriptaddress]{revtex4}

\usepackage{amsmath}
\usepackage{graphicx}
\usepackage{epsfig}

\begin{document}

\title{Loss in hybrid qubit-bus couplings and gates}

\author{Sebastien G.R. Louis}\email{seblouis@nii.ac.jp}
\affiliation{National Institute of Informatics, 2-1-2
Hitotsubashi, Chiyoda-ku, Tokyo 101-8430, Japan}
\affiliation{Department of Informatics, School of Multidisciplinary Sciences,
The Graduate University for Advanced Studies,
2-1-2 Hitotsubashi, Chiyoda-ku, Tokyo 101-8430 Japan}

\author{W.J. Munro}
\affiliation{Hewlett-Packard Laboratories, Filton Road, Stoke
Gifford, Bristol BS34 8QZ, United Kingdom}
\affiliation{National Institute of Informatics, 2-1-2
Hitotsubashi, Chiyoda-ku, Tokyo 101-8430, Japan}

\author{T.P. Spiller}
\affiliation{Hewlett-Packard Laboratories, Filton Road, Stoke
Gifford, Bristol BS34 8QZ, United Kingdom}

\author{Kae Nemoto}
\affiliation{National Institute of Informatics, 2-1-2
Hitotsubashi, Chiyoda-ku, Tokyo 101-8430, Japan}

\begin{abstract}
We provide a characterization and analysis of the effects of dissipation on oscillator assisted (qubus) quantum gates. The effects can be understood and minimized by looking at the dynamics of the signal coherence and its entanglement with the continuous variable probe. Adding loss in between successive interactions we obtain the effective quantum operations, providing a novel approach to loss analysis in such hybrid settings. We find that in the presence of moderate dissipation the gate can operate with a high fidelity. We also show how a simple iteration scheme leads to independent single qubit dephasing, while retaining the conditional phase operation regardless of the amount of loss incurred by the probe.
\end{abstract}


\maketitle

\section{I. introduction}

Quantum information processing can potentially provide a considerable speedup over classical information processing for certain problems \cite{ekert} along with the ability to efficiently simulate physical systems that cannot be done classically \cite{lloyd1,abrams}. In view of this, much work has been done on finding a viable physical implementation of a quantum computer and many different realizations have been proposed \cite{cirac,klm,kane,laddsil,briegel}, some of which explored experimentally on a small scale \cite{vander,sack,pit,walther}. Most large scale architectures rely on protocols enabling the transport of quantum systems or distributed schemes to perform logical gates on isolated qubits. Such schemes make use of `flying' qubits \cite{div,plen-cqed} or more generally of quantum bus concepts \cite{andy,tim}. Decoherence effects on the bus are crucial and being able to overcome them will bring us one step closer to true scalability. 

Most of the results in quantum information theory were developed in a discrete setting, making use of qubits. However even though single qubit operations are not so much of an issue, the level of control required to physically implement entangling gates between individual qubits and measure them within realistic coherence times is tremendous, limiting the experimental realizations. The initial theoretical proposals were rapidly adapted to a continuous variable (CV) setting, where CV quantum information processing  was shown to be possible \cite{sam,loockrev}. CV implementations may be more accessible in some respects, with simple measurements and entangling operations. Despite these advantages, this framework is limited by the nonlinearity available experimentally, making single system operations difficult. Combining the exactness of discrete variables and the robustness of continuous variables is therefore a judicious route to take. The term `hybrid' was first coined by Lloyd \cite{lloyd} to describe quantum information processes featuring both discrete and continuous variables. The ability to switch on and off particular interaction Hamiltonians enables one to simulate interactions and quantum logical gates on discrete systems. Multi-qubit extensions were rapidly undertaken in generalizations \cite{wang,zan}. The initial observation came at the same time as Milburn's proposal to simulate interactions between trapped ions by coupling them to a common vibrational mode \cite{mil}, constituting a direct physical realization of a hybrid quantum computer. This approach was used to entangle up to four ions experimentally \cite{sack}. In recent years other physical implementations have been explored making use of matter qubits \cite{tim}. A hybrid quantum computer could potentially be more  versatile than its strictly discrete counterpart, providing simple algorithms to compute eigenvectors and eigenvalues \cite{lloyd} or implementing Grover's search algorithm in a direct way \cite{wang}. 

A CV bus (qubus) can also be used to achieve quantum non-demolition (QND) measurements and parity gates \cite{qndmil,qndimoto,mun,mun2,seb1}, which for example can be combined with an ancilla qubit to implement a near deterministic CNOT gate \cite{kae}. In these cases the limited strength of the nonlinearity is compensated for by an increase in the bus amplitude, leading to entangling gates based on homodyne measurements functioning at greater success probabilities than in single photon mediated applications. This is particularly relevant to the generation of cluster states \cite{seb1,seb2} and entanglement distribution for repeater applications \cite{loock-rep,ladd}. 

The physical circumstances in which one can envisage a hybrid coupling between a CV and discrete variable system have been extensively investigated. The Jaynes-Cummings model \cite{jay} is very successful at accounting for the interaction of radiation with an atom in a cavity quantum electro-dynamics (CQED) setting. Based on this, one possible realization for the qubus scheme is that of an atomic qubit interacting dispersively with a cavity mode. However superconducting charge qubits are also a good candidate for a physical realization \cite{tim,tim2}, for which a dispersive coupling with a microwave bus mode has already been experimentally demonstrated \cite{wallraff1,wallraff2}. Decoherence effects during such an interaction have been explored in the past, for example in the case of a two-level atom interacting dispersively with an optical mode in a dissipative cavity \cite{nemes,rom,ladd}. This dispersive interaction forms the basis for many qubus schemes and dissipation effects during the interaction on an entangling gate between two qubits can potentially be overcome \cite{bar}. A symmetrization technique to develop resilience to both dissipation and thermal fluctuations was also proposed \cite{cen}. There the authors took advantage of the invariance under time reversal of the action of the gate, noticing that the combination of an interaction sequence with its time reversed version canceled out dissipation effects. Even though they are effective, these methods focus on a pair of qubits within the same cavity or trap and decoherence due to inter-cavity communication remains to be addressed. In addition to this they propose interaction sequences which have to be iterated many times before any significant improvement in gate fidelity can be appreciated. 

The cross-Kerr effect is also well known and used in several different areas of optics. In this context, decoherence effects have been approximated along with their impact on optical quantum information processing using weak nonlinearities \cite{jacob,jacob2}. It was shown that dissipation effects in a two-qubit parity gate can be minimized if one can implement a photon number measurement on the bus. In the present work we extend previous results by solving exactly the master equation during the interaction for arbitrary input states and observing a trade off between the entanglement generated and the required precision in interaction time. We also provide the resulting quantum operations incurred by qubits involved in a CZ gate assuming sequential interactions. These operations are critical when quantum error correction techniques have to be introduced as the system is scaled up.

We note here that direct decoherence effects incured by the qubits, independently from the bus, are being neglected. In so doing we obtain a simplified picture, which can become a good approximation if bus dissipation is the primary source of noise. The eventual optimization of a full architecture will nonetheless need to factor in additional sources of noise.

The paper is structured as follows. In the first section of this paper we derive closed expressions for the effects of dissipation during cross-phase and dispersive Jaynes-Cummings interactions. We follow the entanglement and coherence dynamics of a qubit and the continuous variable as they interact with each other. In section II we carry our attention over to the hybrid gates themselves starting with the conditional displacement gate followed by the CZ gate in section III. By adding dissipation between interactions we obtain the quantum operations undergone by the qubits and discuss gate fidelity. Finally we provide a simple iteration scheme to simplify the operation down to a perfect CZ gate followed by independent single qubit dephasing.

\section{II. interaction loss analysis}

As explained in the introduction we focus here on two types of interactions. The first consists in the cross phase modulation undergone by probe and signal modes as they travel through a Kerr medium. The second type of interaction we consider takes place between an off-resonant optical probe mode and a single atom in a cavity. Their effects are described by interaction Hamiltonians of the form
\begin{equation}
H_{int}=-\hbar\chi a^{\dag}a \hat{\Lambda}
\label{}
\end{equation}
where $a$($a^{\dag}$) are the annihilation (creation) operators acting on the probe mode. In the case of a cross-Kerr interaction, $\hat{\Lambda}=b^{\dag}b$ the number operator acting on the signal mode ($\hat{\Lambda}|n\rangle=\lambda_n|n\rangle=n|n\rangle$, the states $|n\rangle$ corresponding to the Foch state basis) and $\chi$ is proportional to the third order nonlinear susceptibility of the medium. For the CQED setting, the interaction is described by the dispersive limit of the Jaynes-Cummings Hamiltonian \cite{knight} in which $\hat{\Lambda}=Z$ where $Z$ is the Pauli operator acting on the atomic qubit ($Z=|0\rangle\langle 0|-|1\rangle\langle 1|$, $\hat{\Lambda}|n\rangle=\lambda_n|n\rangle=(-1)^n|n\rangle$) and $\chi$ is the atom-light coupling strength \cite{seb1}. In both the cross phase and the CQED settings we have a hybrid interaction between the continuous quadrature variables of the probe field and the discrete degrees of freedom of the subsystem. Initiating the probe in a coherent state $|\alpha\rangle$ and applying the interaction for a time $t$ yields
\begin{equation}
e^{-iH_{int}t/\hbar}\sum_n c_n|n\rangle|\alpha \rangle=\sum_n c_n|n\rangle|\alpha e^{i\lambda_n\chi t} \rangle.
\label{}
\end{equation}

\begin{figure}[!htb]\label{}
\begin{center}\includegraphics[scale=0.4]{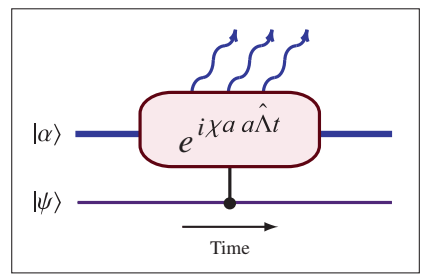}\end{center}
\caption{Loss in the probe mode during the coupling between the discreet system $|\psi\rangle$ and the continuous variable prepared in the coherent state $|\alpha\rangle$.}
\end{figure}

Given these interaction Hamiltonians, we can evaluate the effects of dissipation in the probe mode during the interaction (see Fig. 1) by solving the optical Linblad master equation \cite{gardiner}
\begin{equation}
\frac{\partial \rho(t)}{\partial t}=-\frac{i}{\hbar}\left[H_{int},\rho(t)\right]+\gamma(2a\rho(t) a^{\dag}-a^{\dag}a\rho(t)-\rho(t) a^{\dag}a)
\label{}
\end{equation}
where we have assumed a zero temperature bath (a good approximation in the visible light regime). The damping factor $\gamma$ quantifies the dissipation in the probe mode. If we consider a general input density matrix element $|n\rangle\langle m|\otimes|\alpha\rangle\langle\alpha|$ in which the probe and the signal are disentangled, we can find the equation of motion for this particular element by looking at the operator $\rho_{nm}(t)=\langle n|\rho(t)|m\rangle$. Due to the disentangled form of the initial state we have $\rho_{nm}(0)=|\alpha\rangle\langle\alpha|$ for all $n$ and $m$. The equation of motion for each element is given by
\begin{eqnarray}
\frac{\partial \rho_{nm}(t)}{\partial t}=i\chi \lambda_na^{\dag}a\rho_{nm}(t)+-i\chi \lambda_m \rho_{nm}(t)a^{\dag}a \nonumber \\
+\gamma(2a\rho_{nm}(t) a^{\dag}-a^{\dag}a\rho_{nm}(t)-\rho_{nm}(t) a^{\dag}a).
\label{}
\end{eqnarray}
Following the method used in \cite{nemes}, we use the super-operators $\mathcal{M}(\cdot)=a^{\dag}a(\cdot)$, $\mathcal{P}(\cdot)=(\cdot)a^{\dag}a$ and $\mathcal{J}(\cdot)=a(\cdot)a^{\dag}$ to rewrite the above equation as
\begin{eqnarray}
\frac{\partial \rho_{nm}(t)}{\partial t} &=& \left\{i\chi(\lambda_n\mathcal{M}-\lambda_m\mathcal{P})+\gamma(2\mathcal{J}-\mathcal{M}-\mathcal{P})\right\}\rho_{nm}(t)\nonumber \\
&\equiv& \mathcal{L}_{nm}\rho_{nm}(t),
\label{}
\end{eqnarray}
The formal solution to (5) is then $\rho_{nm}(t)=e^{\mathcal{L}_{nm}t}\rho_{nm}(0)$. The super-operators realize an algebra obeying the commutation relations $\left[\mathcal{J},\mathcal{M}\right]=\left[\mathcal{J},\mathcal{P}\right]=\mathcal{J}$ for which decomposition theorems have been derived \cite{wit}, leading to

\begin{eqnarray}
\mathrm{exp}[\mathcal{L}_{nm}t] &=& \mathrm{exp}\left[\frac{2\gamma(e^{(2\gamma-i(\lambda_n-\lambda_m)\chi)t}-1)}{2\gamma-(\lambda_n-\lambda_m)\chi}\mathcal{J}\right] \nonumber\\
&\times &\mathrm{exp}\left[(i\lambda_n\chi-\gamma)\mathcal{M}t\right] \nonumber\\
&\times &\mathrm{exp}\left[(-i\lambda_m\chi-\gamma)\mathcal{P}t\right].
\label{}
\end{eqnarray}
Now applying this result to our initial element $\rho_{nm}(0)=|\alpha\rangle\langle\alpha|$ we obtain:
\begin{eqnarray}
\rho_{nm}(t)&=&\mathrm{exp}[-|\alpha|^2\{1-e^{-2\gamma t}-\frac{1-e^{(-2\gamma+i(\lambda_n-\lambda_m)\chi)t}}{1-i(\lambda_n-\lambda_m)\chi/2\gamma}\}] \nonumber \\
&\times &|\alpha e^{(-\gamma+i\lambda_n\chi) t}\rangle\langle\alpha e^{(-\gamma+i\lambda_m\chi) t}|.
\label{}
\end{eqnarray}
The coefficient derived above is the closed expression for the `coherence parameter' given in \cite{jacob}. If we denote this coefficient by $\zeta_{nm}$ then we have $\zeta_{nm}=\zeta^{\ast}_{mn}$ and so $|\zeta_{nm}|=|\zeta_{mn}|$ as well as $\zeta_{nn}=1$. We also quickly notice from (7) that this coherence parameter does not tend to 0 as $t$ tends to infinity for a fixed $\alpha$. We have

\begin{equation}
|\zeta_{nm}|_{t\rightarrow\infty}=\mathrm{exp}\left[-|\alpha|^2\frac{(\lambda_n-\lambda_m)^2}{4(\gamma/\chi)^2+(\lambda_n-\lambda_m)^2}\right].
\label{}
\end{equation}
One way of understanding this is that the probe undergoes loss as it couples to the signal, thus reducing the coherence in the signal. But eventually the probe returns to the vacuum state (the time it takes depends on the initial amplitude and the damping factor), disentangling itself and leaving some coherence in the signal. However the larger the amplitude, the larger the effective interaction time, the less coherence remains in the signal.

As time progresses, the process can be viewed as a unitary operation between the signal and the damped probe in addition to a dephasing effect on the signal. To view this more clearly, let us write the output density matrix using $\hat{\Lambda}=Z$, $\theta=\chi t$ and defining $z_n\equiv(-1)^n$,
\begin{eqnarray}
\rho(t)&=&\sum_{n,m=0,1}c_{nm}\zeta_{nm}|n\rangle\langle m|\nonumber\\
&\otimes &|\alpha e^{-\gamma t+iz_n\theta}\rangle\langle\alpha e^{-\gamma t+iz_m\theta}|.
\label{} 
\end{eqnarray}
Writing $\zeta_{nm}=e^{f_{nm}}$, the factor $e^{\mathrm{Re}[f_{nm}]}$ characterizes the decoherence and takes the form $e^{-\epsilon (1-z_nz_m)}$ (see the Appendix for the full expression). Applying this type of operation to a qubit density matrix $\rho=\sum_{n,m=0,1}c_{nm}|n\rangle\langle m|$ yields directly the phase flip channel \cite{niel}
\begin{eqnarray}
e^{\mathrm{Re}[f_{nm}]}\rho &=& e^{-\epsilon(1-z_nz_m)}\rho \nonumber\\
&=& e^{-\epsilon}(\mathrm{cosh}\epsilon+z_nz_m\mathrm{sinh}\epsilon)\rho \nonumber\\
&=& \frac{1+e^{-2\epsilon}}{2}\rho+\frac{1-e^{-2\epsilon}}{2}Z\rho Z.
\label{}
\end{eqnarray}
The additional phase $e^{\mathrm{Im}[f_{nm}]}$ acquired in the process (see Appendix) is known and can be corrected for if needed; it is not an intrinsic source of noise.

One expects this issue of coherence to be intimately linked to the entanglement shared between the signal and the probe systems. In order to observe the dynamics of this entanglement we restrict the signal to being a qubit and will continue to use $\hat{\Lambda}=Z$. We also take $\alpha$ real for simplicity. Equation (7) provides us with a time dependent density matrix and having our input signal in the state $(|0\rangle+|1\rangle)/\sqrt{2}$, it reads
\begin{eqnarray}
\rho(t)&=&\frac{1}{2}\{|0\rangle\langle 0|\otimes|\alpha_0\rangle\langle \alpha_0|+\zeta_{01}|0\rangle\langle 1|\otimes|\alpha_0\rangle\langle \alpha_1| \nonumber \\
&+& \zeta_{10}|1\rangle\langle 0|\otimes|\alpha_1\rangle\langle \alpha_0|+|1\rangle\langle 1|\otimes|\alpha_1\rangle\langle \alpha_1|\},
\label{}
\end{eqnarray}
with $\alpha_0=\alpha e^{(-\gamma+i\chi) t}$ and $\alpha_1=\alpha e^{(-\gamma-i\chi) t}$. The entanglement being invariant under local unitary operations we allow ourselves for simplicity to apply the conditional phase $|0\rangle\langle 0|+e^{i\mathrm{Im}[\alpha_0\alpha_1^{\ast}]}|1\rangle\langle 1|$ on the qubit. Then we redefine the bus probe states as $|\alpha_0\rangle$ and $|\alpha_1'\rangle=e^{i\mathrm{Im}[\alpha_0\alpha_1^{\ast}]}|\alpha_1\rangle$ so that the overlap  between the two is real: $\langle \alpha_0|\alpha_1'\rangle=|\langle \alpha_0|\alpha_1\rangle|$. This allows us to express them in an orthogonal basis $\{|x\rangle,|y\rangle\}$ as \cite{norb}
\begin{eqnarray}
&&|\alpha_0\rangle=a|x\rangle+b|y\rangle,\nonumber\\
&&|\alpha_1'\rangle=a|x\rangle-b|y\rangle.
\label{}
\end{eqnarray}
Taking $a$ and $b$ real without loss of generality, normalization leads to
\begin{equation}
a=\sqrt{\frac{1+\delta}{2}},\quad b=\sqrt{\frac{1-\delta}{2}},
\label{}
\end{equation}
with $\delta=e^{-\alpha^2e^{-2\gamma t}(1-\mathrm{cos}(2\chi t))}$. At this point we can write our locally equivalent density matrix in the orthonormal basis $\{|0\rangle|x\rangle,|0\rangle|y\rangle,|1\rangle|x\rangle,|1\rangle|y\rangle\}$ as follows
\begin{equation}
\rho(t)=\frac{1}{2}\left(\begin{array}{cccc}
a^2 & ab & \zeta_{01}a^2 & -\zeta_{01}ab\\
ab & b^2 & \zeta_{01}ab & -\zeta_{01}b^2\\
\zeta_{10}a^2 & \zeta_{10}ab & a^2 & -ab \\
-\zeta_{10}ab & -\zeta_{10}b^2 & -ab & b^2\end{array} \right).
\end{equation}
We have now managed to express the qubit and continuous variable composite state in the form of a two qubit state. Given the resulting two-qubit density matrix, there are several entanglement measures to choose from, including the logarithmic negativity and the relative entropy of entanglement \cite{plen-ent-rev}. Here we will work with the concurrence as defined by Wooters \cite{woot,woot2}, which we plot as a function of the scaled time $\chi t$ for particular choices of parameters $\alpha$ and $\gamma/\chi$. As the qubit and the field initially start in a product state, and eventually for large times should return to a product state when the probe field doesn't contain anymore photons, we would expect the entanglement to peak at some point in time. 

\begin{figure}[!htb]\label{Fig2}
\begin{center}\includegraphics[scale=0.5]{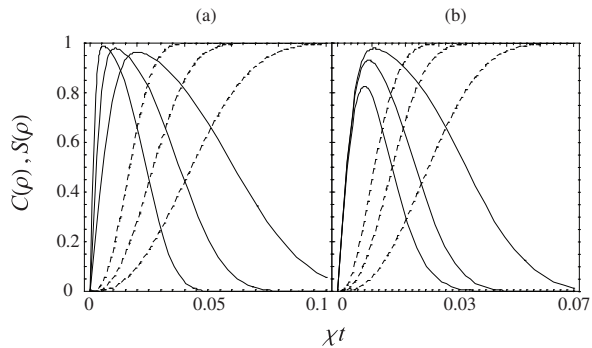}\end{center}
\caption{Plots of the concurrence $C(\rho)$ (solid) and the von Neumann entropy $S(\rho)=-\mathrm{tr}(\rho \mathrm{log} \rho)$ (dashed) of the combined state of the continuous variable mode and qubit as a function of the scaled time $\chi t$. (a) From left to right $\alpha$=200,100 and 50 with a fixed ratio of damping rate to nonlinearity $\gamma/\chi=1$. (b) The amplitude $\alpha$ is fixed to 100 and from lowest to highest peaking curves $\gamma/\chi$=1, 7 and 21.}
\end{figure}
This is verified in the plots of Fig. 2. In Fig. 2(a) we can observe how the amplitude of the field $\alpha$ affects the behavior of entanglement in time. As expected the larger $\alpha$ is, the larger the maximum entanglement. This is simply explained by the fact that the separation between the possible states of the field in phase space increases with $\alpha$, for a same interaction strength, thus making them more distinguishable. For large $\alpha$ the maximum concurrence tends naturally to 1, however the peaking of the entanglement also becomes sharper. It is quickly generated, but also quickly destroyed, as illustrated by the von Neumann entropy $S(\rho)$ characterizing the decoherence which is a function of the squared distance in phase space between the two field states. Fig. 2(b) shows us how the maximum achievable entanglement depends on the ratio $\gamma/\chi$, as did the limit of the coherence parameter in the previous section. The larger the relative damping $\gamma$, the lower and the quicker the entanglement peaks in time. In both plots $S(\rho)$ tends to 1, meaning the qubit is left in a maximally mixed state, disentangled from the probe. However this is not always the case and in general the smaller the ratio $\chi/\gamma$ is, the lower the final entropy of the qubit becomes.

In view of a QND measurement on a single qubit \cite{mil2,imo}, only the entanglement with the probe needs to be taken into account, as decoherence in the process will not affect the measurement statistics. However, when the application becomes cat state generation or multiqubit gates \cite{jacob,jacob2}, decoherence becomes a crucial issue which has to be weighted against the entanglement. In such applications one wishes to produce coherent superpositions of single or multiple quantum systems. Thus it is important to view the behavior of the entropy of the combined state at the time at which the entanglement is maximized. This behavior is illustrated in Fig. 3(a), showing the expected limiting behavior of the maximum entanglement as $\alpha$ increases. In Fig. 3(b) the corresponding entropy of the combined outgoing state is seen to decrease asymptotically for all choices of the parameter $\gamma/\chi$. In consequence one can simply reduce the amount of decoherence by increasing the strength of the probe. This is in part due to the fact that the interaction time becomes shorter, reducing the effective decoherence time. 

The success of such an approach to minimize the decoherence will then depend on the loss incurred in between interactions. The reason being that the larger the amplitude $\alpha$ is, the larger the amount of dephasing incured by the qubits coupled to the probe mode during these time intervals will be. This will become clear in the next section. In consequence we observe a similar trade-off of as that encountered in schemes such as the hybrid quantum repeater proposed in \cite{ladd}. If the transit time and conditions are appropriate, the approach is indeed effective. For example taking $\gamma/\chi=5$ and a reasonable amplitude $\alpha=10^4$ we obtain a maximum concurrence of 0.998 for a von Neumann entropy of $10^{-2}$.

\begin{figure}[!htb]\label{Fig3}
\begin{center}\includegraphics[scale=0.5]{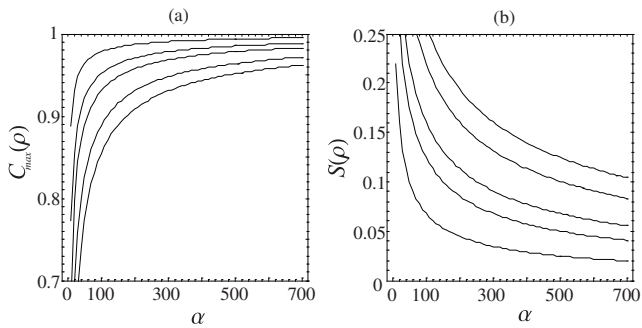}\end{center}
\caption{(a) The maximum concurrence as a function of the amplitude $\alpha$ of the probe with $\gamma/\chi=1$, 3, 5, 10 and 15 from top to bottom. (b) The von Neumann entropy of the combined state at the entanglement peaking time with the same values for $\gamma/\chi$ in decreasing order from top to bottom.}
\end{figure}

So the higher the entanglement we want to measure or couple out of the cavity if we are dealing with cavity QED systems, the larger the probe amplitude and the more precise the timing of the interaction will have to be.  Clearly, these issues of coherence and entanglement will have to be combined in order to optimize quantum gates in which different qubits interact with the same probe mode.

\section{III. The conditional displacement}

Many qubus gate proposals \cite{loock-rep,tim,bar,seb1} rely directly on the dispersive Jaynes-Cummings interaction. In those schemes different qubits in separate cavities interact sequentially with the same probe mode and subsequent homodyne measurements can project the qubits to entangled states in a heralded fashion. Alternatively photon number resolved detection can be used, potentially leading to deterministic gates. Dissipation in the probe mode during such gates will reduce the fidelity of the post-selected entangled states and has been investigated in some detail \cite{ladd}. An interesting trade-off arises between the distinguishability of the measurement outcomes and the amount of dephasing incured by the qubits.

More in line with the initial hybrid proposals \cite{mil,wang,zan}, measurement-free quantum gates can be implemented with the use of what we will refer to as the \textit{conditional displacement} interaction. A qubit conditionally displaces a continuous variable bus with the following operation: $D(\beta Z)=\mathrm{exp}[(\beta \hat{a}^{\dag}-\beta ^{\ast}a)Z]$ where $\beta$ is the amount by which the field is displaced. Thus we can see that after the operation, the position of the probe in phase space is dependent on the state of the qubit (they have become entangled). 

There are different ways of arriving at such a coupling between a qubit and a continuous variable. The physical system may for example be naturally described by an interaction Hamiltonian directly producing the conditional displacement as defined above. A flux qubit in a cavity for example could displace the cavity mode in such a way \cite{blais}. In cavity QED systems, a particularly shaped pulse entering and leaking out of the cavity could also provide us with a very similar interaction Hamiltonian \cite{billprep}. However, based on the dispersive Jaynes-Cummings interaction, it was shown lately that a conditional displacement as defined above can be exactly simulated by using unconditional displacements \cite{loock}. Defining the \textit{conditional rotation} operation encountered in the previous section as $R(\theta Z)=\mathrm{exp}[i\theta a^{\dag}a Z]$, a possible sequence of interactions is the following

\begin{equation}
D(\alpha \mathrm{cos}\theta)R(-\theta Z)D(-2\alpha)R(\theta Z)D(\alpha\mathrm{cos}\theta)=D(2i\alpha\mathrm{sin}\theta Z),
\label{}
\end{equation}
with $\alpha$ real (see Fig. 4). Surprisingly all the phases cancel out to give us an exact conditional displacement, independent of the initial probe position in phase space. Making $\alpha$ pure imaginary will lead to a displacement in the orthogonal (real) direction. Consecutive displacements induce phases as $D(\beta)D(\alpha)=e^{i \mathrm{Im}(\alpha^{\ast}\beta)}D(\alpha+\beta)$ and consequently $D^{\dag}(\alpha)D^{\dag}(\beta)=e^{-i \mathrm{Im}(\alpha^{\ast}\beta)}D^{\dag}(\alpha+\beta)$ as $D^{\dag}(\delta)=D(-\delta)$. Displacements and rotations acting on the density matrix we will denote as $\mathcal{D}(\alpha)\rho=D(\alpha)\rho D^{\dag}(\alpha) $ and $\mathcal{R}(\theta)\rho=R(\theta)\rho R^{\dag}(\theta)$.

\begin{figure}[!htb]\label{Fig4}
\begin{center}\includegraphics[scale=0.4]{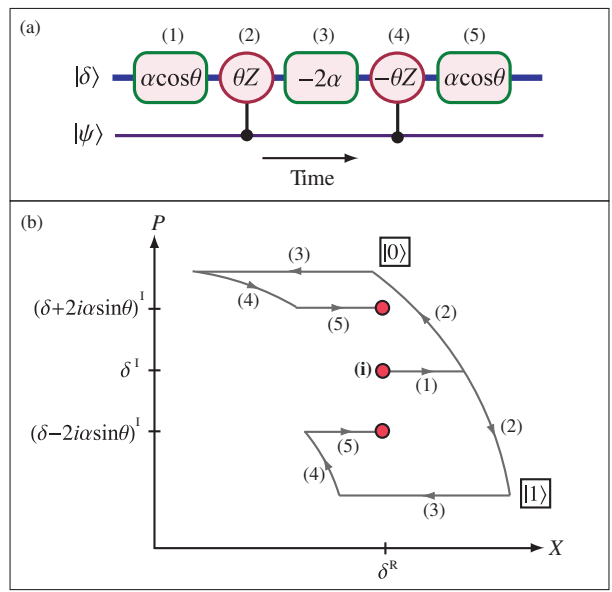}\end{center}
\caption{(a) The interaction sequence for the the simulation of the conditional displacement gate. (b) The two possible trajectories of the probe in phase space, the upper and lower paths corresponding to the qubit states $|0\rangle$ and $|1\rangle$ respectively. Here as in the text, we assume $\alpha$ to be real. The `R' and `I' exponents denote the real and imaginary parts of the probe amplitude $\delta$ at each time step.}
\end{figure}

In order to characterize loss in an interaction sequence, we will introduce dissipation in the probe mode between and during each interaction.  Dissipation during the interactions is due to loss in the nonlinear material or cavity system used to to mediate the interaction, whereas dissipation in between each interaction is due to fiber loss, mode mismatch and other effects. Consequently a different loss parameter should arise. However, both types of loss result in a dephasing of the qubit(s) and their effects can thus be combined into a single loss parameter. We will also assume the amount of loss is the same in each segment. Dissipation affects a coherent state matrix element as $\mathcal{L}|\alpha\rangle\langle\beta|\rightarrow \langle \beta|\alpha\rangle ^{\eta}|\dot{\alpha}\rangle\langle\dot{\beta}|$ with $\eta=1-e^{-2\gamma t}$ \cite{walls}. We fix the amount of loss between each interaction $l=\gamma t$ and represent the attenuated coherent state by $\dot{\alpha}=\alpha e^{-l}$ such that the number of dots will determine the number of attenuations i.e. $\ddot{\alpha}=\alpha e^{-2l}$. The quantum operations on the qubits will be obtained by calculating these state-dependent overlaps.

Now let us consider the effects of dissipation in the whole interaction sequence (15). For generality we will keep the amplitudes of the three displacements as free real variables $\alpha_1$, $\alpha_2$ and $\alpha_3$. The first step in the sequence is a displacement so taking our probe initially in the vacuum state we have
\begin{eqnarray}
\mathcal{D}(\alpha_1)\rho &=& \sum_{a,a'=0,1}c_{aa'}|a\rangle\langle a'|\otimes D(\alpha_1)|0\rangle\langle 0|D^{\dag}(\alpha_1) \nonumber \\
&=& \sum_{a,a'=0,1}c_{aa'}|a\rangle\langle a'|\otimes|\alpha_1\rangle\langle \alpha_1|,
\label{}
\end{eqnarray}
where $\{|a\rangle,a=0,1\}$ represents the basis states of the qubit and the state of the probe mode is kept to the right. Loss in the probe mode at this point will decrease the amplitude of the coherent state to $|\dot{\alpha}_1\rangle$, without affecting the qubit. Proceeding to the second step which is a conditional rotation we have
\begin{equation}
\mathcal{R}(\theta Z).\mathcal{L}.\mathcal{D}(\alpha_1)\rho=\sum_{a,a'=0,1}c_{aa'}|a\rangle\langle a'|\otimes|\dot{\alpha}_1 e^{i\theta z_a}\rangle\langle \dot{\alpha}_1e^{i\theta z_{a'}}|.
\label{}
\end{equation}
Now we introduce loss in the probe mode, leading to
\begin{eqnarray}
\mathcal{L}.\mathcal{R}(\theta Z).\mathcal{D}(\alpha_1)\rho &=& \sum_{a,a'=0,1}c_{aa'}|a\rangle\langle a'|\otimes|\ddot{\alpha}_1 e^{i\theta z_a}\rangle\langle \ddot{\alpha}_1e^{i\theta z_{a'}}| \nonumber \\
&\times& \langle \dot{\alpha}_1e^{i\theta z_{a'}}|\dot{\alpha}_1 e^{i\theta z_a}\rangle^{\eta}.
\label{}
\end{eqnarray}
Continuing in this fashion and completing the sequence
\begin{eqnarray}
&&\mathcal{C}_a\rho=\mathcal{D}(\alpha_3).\mathcal{L}.\mathcal{R}(-\theta Z).\mathcal{L}.\mathcal{D}(\alpha_2).\mathcal{L}.\mathcal{R}(\theta Z).\mathcal{L}.\mathcal{D}(\alpha_1)\rho \nonumber \\
\nonumber\\
&&=\sum_{a,a'=0,1}c_{aa'}|a\rangle\langle a'|\otimes|\ddddot{\alpha}_1+\ddot{\alpha}_2 e^{-i\theta z_a}+\alpha_3\rangle\langle z_a\rightarrow z_{a'}|\nonumber \\
&&\times \mathrm{exp}[i\mathrm{sin}\theta( \ddot{\alpha}_2\alpha_3-\ddot{\alpha}_1\alpha_2)(z_a-z_{a'})] \times (\xi_1\xi_2\xi_3)^{\eta},
\label{}
\end{eqnarray}
with the $\xi$'s representing the three loss terms (overlaps). The notation $z_a\rightarrow z_{a'}$ means the contents of the bra are the same as in the previous ket replacing $z_a$ with $z_{a'}$. In order to simulate a conditional displacement (with dephasing on the qubit), we require the state of the probe mode to be of the form $|\gamma z_a\rangle\langle \gamma z_{a'}|$. This is achieved by setting $\ddddot{\alpha}_1+\ddot{\alpha}_2\mathrm{cos}\theta+\alpha_3=0$. Combining the loss terms we obtain

\begin{equation}
\xi_1\xi_2\xi_3=\mathrm{exp}[-S(1-z_az_{a'})]\times\mathrm{exp}[iT(z_a-z_{a'})],
\label{}
\end{equation}
with $T=\mathrm{sin}\theta(\ddot{\alpha}_1\alpha_2+\dddot{\alpha}_1\dot{\alpha}_2+(\dot{\alpha}_1^2+\ddot{\alpha}_1^2+\dot{\alpha}_2^2)\mathrm{cos}\theta)$ and $S=\mathrm{sin}^2\theta(\dot{\alpha}_1^2+\ddot{\alpha}_1^2+\dot{\alpha}_2^2)$. Here as in the evaluation of the coherence parameter (7) in section I, the exponent can be separated into real and imaginary parts. The former is characterized by $S$, representing the amount of dephasing incurred by the qubit, which can be decomposed into a phase flip channel as in (10). The latter constitutes the known phase acquired by the qubit in the process, characterized by $T$. Interestingly, this overall conditional phase can be tuned at will by adapting the amplitudes of the displacements, leading to the exact simulation of a conditional displacement. In other words, in this sequence we can limit the effects of dissipation to a dephasing on the qubit, producing nonetheless the correct combined output state. Previously also, loss during the interaction had led to single qubit dephasing effects which can be factored in here. We now move on to examine a two qubit gate in the presence of probe loss.

\section{IV. The CZ gate}

Based on four of these conditional displacements induced by a pair of qubits on the same probe mode, a gate locally equivalent to the CZ gate can be built. An interaction sequence leading to this unitary operation is $\hat{U}=D(-\beta_bZ_b)D(-\beta_aZ_a)D(\beta_bZ_b)D(\beta_aZ_a)=e^{2iIm\left\{\beta_a^*\beta_b\right\}Z_aZ_b}$ with $\beta_a^{\ast}\beta_b=i\pi/8$ where $Z_a$ and $Z_b$ act on qubits $a$ and $b$ respectively. Here we use the same method as before, introducing dissipation between each interaction, as illustrated in Fig. 5. We will assume the probe starts off in the vacuum state, but the same result is obtained if we initialize it in any coherent state. The initial density matrix reads

\begin{equation}
\rho_{ab}=\sum_{a,a',b,b'=0,1}c_{aa'bb'}|ab\rangle\langle a'b'|\otimes|0\rangle\langle0|.
\label{}
\end{equation}
At this point it is worth noting that even though the probe undergoes amplitude damping, it can nonetheless be perfectly disentangled from the qubits coupled to it whatever the amplitude of the coherent state it starts off in (it may not be the case for other optical states). This is done by tuning the second (opposite) conditional displacement, in function of the known loss parameter $l$ into which loss during the interaction has also been factored. Thus the amplitude of the second conditional displacement will be reduced by a factor of $e^{-2l}$, the damping undergone by the probe since the last coupling. Resolving the whole gate sequence (see Fig. 5) and choosing $\beta_a$ and $\beta_b$ to be real we obtain

\begin{figure}[!htb]\label{Fig4}
\begin{center}\includegraphics[scale=0.4]{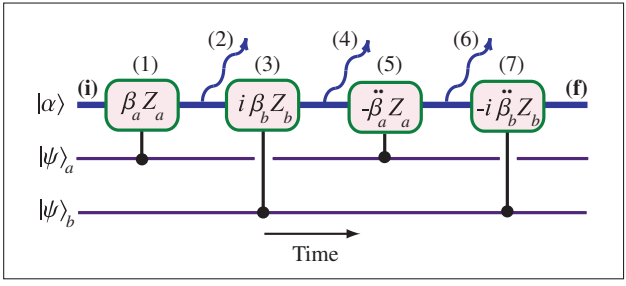}\end{center}
\caption{The interaction sequence for the CZ gate with loss. Between each one of the four conditional displacements (1,3,5,7) the probe undergoes dissipation (2,4,6).}
\end{figure}

\begin{eqnarray}
&&\mathcal{D}(-i\ddot{\beta}_bZ_b).\mathcal{L}.\mathcal{D}(-\ddot{\beta}_aZ_a).\mathcal{L}.\mathcal{D}(i\beta_bZ_b).\mathcal{L}.\mathcal{D}(\beta_aZ_a)\rho=\nonumber \\
\nonumber \\
&&\sum_{a,a',b,b'=0,1}c_{aa'bb'}|ab\rangle\langle a'b'|\otimes|0\rangle\langle 0|\times\left(\xi_1\xi_2\xi_3\right)^{\eta}\nonumber\\
&&\times\mathrm{exp}[i(\dot{\beta}_a\beta_b+\ddot{\beta}_a\dot{\beta}_b)(z_az_b-z_{a'}z_{b'})].
\label{}
\end{eqnarray}
First we notice the geometrical phase represented by the last term. This is precisely the form of a two-qubit conditional phase having been applied to the density matrix. Ignoring the other terms, if we can set $\dot{\beta}_a\beta_b+\ddot{\beta}_a\dot{\beta}_b=\pi/4$ then we have simulated a CZ gate. However the dephasing effects are included in the three $\xi$ overlaps. The first and third lead to single qubit dephasing

\begin{eqnarray}
\xi_1=\langle \beta_a z_{a'}|\beta_a z_a\rangle=\mathrm{exp}[-\beta_a^2(1-z_az_{a'})],\nonumber\\
\nonumber\\
\xi_3=\langle i\dot{\beta}_bz_{b'}|i\dot{\beta}_b z_b\rangle=\mathrm{exp}[-\dot{\beta}_b^2(1-z_bz_{b'})],
\label{}
\end{eqnarray}
while the second overlap corresponds to loss in the probe mode when it holds information on both qubits
\begin{eqnarray}
&&\xi_2=\langle \dot{\beta}_a z_{a'}+i\beta_bz_{b'}|\dot{\beta}_a z_a+i\beta_bz_b\rangle\nonumber\\
&&=\mathrm{exp}[-(\dot{\beta}_a^2+\beta_b^2)+(\dot{\beta}_az_a+i\beta_bz_b)(\dot{\beta}_az_{a'}-i\beta_bz_{b'})].\nonumber\\
\label{}
\end{eqnarray}
In order to be able to express the resulting quantum operation in a closed form, we first arrange the terms in the exponential such that when writing the expansion we obtain an accessible closed algebra. Symmetrizing the terms we have

\begin{figure}[!htb]\label{Fig5}
\begin{center}\includegraphics[scale=0.41]{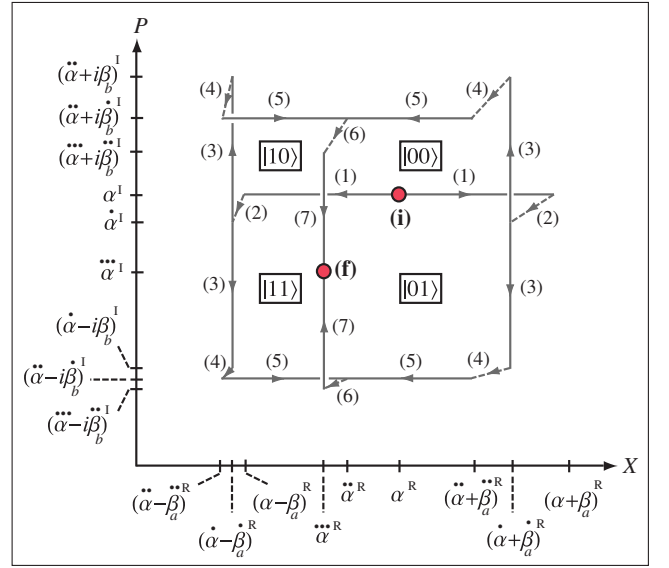}\end{center}
\caption{The trajectories of the probe state in phase space. All four paths start and end at the same amplitudes $\alpha$ and $\dddot{\alpha}$ respectively, insuring that the probe disentangles from the qubits.}
\end{figure}

\begin{equation}
\xi_2^{\eta}=\mathrm{exp}[-x_0+x_1z_az_{a'}+x_2 z_bz_{b'}+x_3(z_a+iz_b)(z_{a'}-i z_{b'})],
\label{}
\end{equation}
with $x_0=\eta(\dot{\beta}_a^2+\beta_b^2)$, $x_1=\eta\dot{\beta}_a(\dot{\beta}_a-\beta_b)$, $x_2=\eta\beta_b(\beta_b-\dot{\beta}_a)$ and $x_3=\eta\dot{\beta}_a\beta_b$. Focusing first on the $x_3$ term, we can write out the expansion of that particular exponential acting on the density matrix as
\begin{eqnarray}
\sum_{n=0}^{\infty}\frac{(x_3)^n(Z_a+iZ_b)^n\rho(Z_a-iZ_b)^n}{n!}.
\end{eqnarray}
Now we define the two-qubit operators
\begin{eqnarray}
&&J=Z_a+iZ_b, \nonumber\\
&&K=Z_aZ_b.
\label{}
\end{eqnarray}
Then grouping the terms in the expansion we obtain the following unnormalized operation
\begin{equation}
e^{x_3(z_a+iz_b)(z_{a'}-i z_{b'})}\rho=c_+\rho+c_- K\rho K+s_- J \rho J^{\dag}+s_+ J^{\dag} \rho J,
\label{}
\end{equation}
with
\begin{eqnarray}
c_{\pm} &=& (\mathrm{cosh}2x_3\pm\mathrm{cos}2x_3)/2,\nonumber\\
s_{\pm} &=& (\mathrm{sinh}2x_3\pm\mathrm{sin}2x_3)/4.
\label{}
\end{eqnarray}
Now we have to factor in the other terms $x_0$, $x_1$ and $x_2$ of $\xi_2^{\eta}$. Further identifying
\begin{eqnarray}
e_0 &=& \mathrm{cosh}x_1\mathrm{cosh}x_2,\quad e_1 = \mathrm{cosh}x_1\mathrm{sinh}x_2,\nonumber\\
e_2 &=& \mathrm{sinh}x_1\mathrm{cosh}x_2,\quad e_3 = \mathrm{sinh}x_1\mathrm{sinh}x_2,
\label{}
\end{eqnarray}
and $K'=i+K$ we obtain the final normalized operation
\begin{eqnarray}
\xi_2^{\eta}\rho &=& e^{-x_0}\{ (c_+e_0+c_-e_3)\rho+(c_+e_2+c_-e_1)Z_a\rho Z_a\nonumber\\
&+& (c_+e_1+c_-e_2)Z_b\rho Z_b+(c_+e_3+c_-e_0)K\rho K\nonumber\\
&+& (s_+e_1+s_-e_2)K'\rho K'^{\dag}+(s_+e_2+s_-e_1)K'^{\dag}\rho K'\nonumber\\
&+& (s_+e_3+s_-e_0)J\rho J^{\dag}+(s_+e_0+s_-e_3)J^{\dag}\rho J\},
\label{}
\end{eqnarray}
on the two-qubit state. Setting $\beta_b=\dot{\beta}_a=\beta$ removes the single qubit terms $Z_a$ and $Z_b$ and also the $K'$ terms, yielding the operation
\begin{equation}
\xi_2^{\eta}\rho=e^{-2\eta\beta^2}(c_+\rho+c_-K\rho K+s_-J\rho J^{\dag}+s_+J^{\dag}\rho J) 
\label{}
\end{equation}
with $x_3=\eta\beta^2$. Focusing on the low loss regime, we can assume a small $\eta$ between each interaction. Truncating to second order in $\eta$ the expansion (26) leading to the above operation we have
\begin{eqnarray}
\rho & \rightarrow & \rho+\eta\beta^2(Z_a+iZ_b) \rho (Z_a-iZ_b)\nonumber \\
&=& \rho+\eta\beta^2 Z_a\rho Z_a+\eta\beta^2Z_b\rho Z_b \nonumber \\
&+& i\eta\beta^2(Z_b\rho Z_a-Z_a\rho Z_b).
\label{}
\end{eqnarray}
Let us assume we are in a quantum error correction (QEC) setting where ancilla qubits are being used. Once the ancilla systems have undergone projective measurements, syndromes are extracted for each logical qubit, indicating whether or not it has been subject to a $Z$ error. The last two terms however will lead to cross terms of syndrome states which cannot be observed in the measurement process and thus they are removed from the resulting density matrix \cite{dev}. This leaves us with single qubit errors on $a$ and $b$ with equal probability (they constitute the observable part of the map). So for small loss we are only observing single qubit errors throughout the gate (at each one of the three dissipation stages) which can be corrected for via QEC. 

Correlated errors represented by the operator $K$ appearing in higher order terms are quantified by the coefficient $c_-$ in the normalized operation (32). In Fig. 7(a) we can see how this part scales with loss in comparison to uncorrelated errors. The general quantum operations we have obtained are of great importance for error correction. They provide us with the error syndromes and their associated probabilities which in turn can be directly fed into fault tolerance calculations. This will ultimately allow us to compare the qubus scheme with other proposed implementations of quantum information processing.

\begin{figure}[!htb]\label{Fig3}
\begin{center}\includegraphics[scale=0.5]{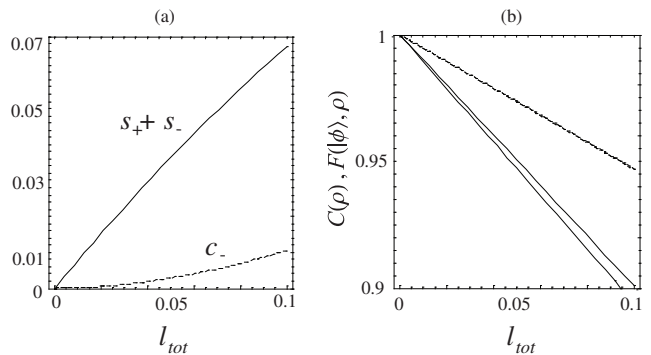}\end{center}
\caption{(a) The scaling of the normalized correlated ($c_-$) and uncorrelated errors ($s_++s_-$) against the loss $l_{tot}$ as defined in the text. Dissipation occurs three times, once between each interaction, in equal amounts. (b) The fidelity (dashed) $F(|\phi\rangle,\rho)=\langle \phi|\rho|\phi\rangle$ of the two qubit output state, where $|\phi\rangle=e^{i\pi Z_1Z_2/4}|+\rangle_1|+\rangle_2$ constitutes the ideal output state and the concurrence (solid), both for the single and the iterated sequences. We observe the same fidelity for both of them while the outputted concurrence is lower for the iterated sequence.}
\end{figure}

To appreciate the effect of loss on the whole gate we look at the fidelity of the output state with regards to the ideal output, for a two qubit input state $|+\rangle|+\rangle$. Such an equally weighted superposition input state provides a good general indication as to the gates performance in addition to the fact that the ideal output is a useful resource, locally equivalent to a Bell state or two-qubit cluster state. A plot of the fidelity against the relative intensity decrease of the probe through dissipation defined as $l_{tot}=1-(\mathrm{exp}[-3l])^2$ is shown in Fig. 7(b). This can be understood in that if the probe is initiated in a coherent state with amplitude $\alpha$, then at the end of the gate it disentangles from the qubits and is left with an amplitude $\alpha e^{-3l}$. In computing the fidelity we use 
\begin{equation}
\beta_a=\beta_b=\frac{1}{2}\sqrt{\frac{\pi}{e^{-l}+e^{-3l}}}
\label{}
\end{equation}
so as to make the phase represented by the last term in (22) that of an ideal CZ gate. We find that even for an intensity decrease of up to 80\% in the probe mode (corresponding to 8 dB loss in total), the output fidelity remains above 0.5, allowing for purification. In the moderate loss regime, both the fidelity and the entanglement remain high; for example taking $l_{tot}=0.05$ corresponding to a decrease of $5\%$ in probe intensity (0.22 dB loss) results in an output with $F\sim 0.97$ and a concurrence of $\sim 0.95$. 

Finally we now show how a simple repetition scheme can significantly simplify full gate operation. The first point to notice is similar in spirit to the observation made in \cite{cen} that the ideal (loss-free) operation is invariant under time reversal. That is if we reverse the order of the interactions, we obtain the same conditional geometric phase. However in the case of separate cavities, a time reversed iteration does not help fight decoherence for a coherent state probe. The reason for this is that the single qubit error linked to the transfer between each cavity scales in the same way as the geometrical phase (of order $2\beta^2$). The observation we make here is that the gate is also invariant under a swapping of the displacement directions. That is the same geometrical phase is obtained, again in the loss free case, if now qubit $a$ conditionally displaces the probe in the imaginary direction in (22) and qubit $b$ displaces it in the real direction. Let us denote the two different sequences in a dissipative setting as

\begin{eqnarray}
\mathcal{S}&=&\mathcal{D}(-i\ddot{\beta}_bZ_b).\mathcal{L}.\mathcal{D}(-\ddot{\beta}_aZ_a).\mathcal{L}.\mathcal{D}(i\beta_bZ_b).\mathcal{L}.\mathcal{D}(\beta_aZ_a)\nonumber\\
\nonumber\\
\tilde{\mathcal{S}}&=&\mathcal{D}(-\ddot{\beta}_bZ_b).\mathcal{L}.\mathcal{D}(-i\ddot{\beta}_aZ_a).\mathcal{L}.\mathcal{D}(\beta_bZ_b).\mathcal{L}.\mathcal{D}(i\beta_aZ_a).\nonumber\\
\label{seq}
\end{eqnarray}
We have

\begin{eqnarray}
&&\mathcal{S}=e^{i\kappa(z_az_b-z_{a'}z_{b'})}(\xi_1\xi_2\xi_3)^{\eta}\mathcal{D}(0)\nonumber\\
\nonumber\\
&&\tilde{\mathcal{S}}=e^{i\kappa(z_az_b-z_{a'}z_{b'})}(\xi_1\tilde{\xi}_2\xi_3)^{\eta}\mathcal{D}(0),
\label{effect}
\end{eqnarray}
with $\kappa=\dot{\beta}_a\beta_b+\ddot{\beta}_a\dot{\beta}_b$. Now we notice that the effects of the two sequences differ only in the central overlap terms $\xi_2$ and $\tilde{\xi}_2$ which contain the multiple correlated error terms. It is straightforward to see that their combined effect yields

\begin{eqnarray}
\tilde{\xi}_2\xi_2&=&\langle i\dot{\beta}_a z_{a'}+\beta_bz_{b'}|i\dot{\beta}_a z_a+\beta_bz_b\rangle\nonumber\\
&\times&\langle \dot{\beta}_a z_{a'}+i\beta_bz_{b'}|\dot{\beta}_a z_a+i\beta_bz_b\rangle\nonumber\\
&=& e^{-2\beta_b^2(1-z_bz_{b'})}e^{-2\dot{\beta}_a^2(1-z_az_{a'})},
\label{combief}
\end{eqnarray}
which are just single qubit phase flip channels. Thus the combination of the two sequences gives the operation

\begin{eqnarray}
\tilde{\mathcal{S}}\mathcal{S}&=&e^{2i\kappa(z_az_b-z_{a'}z_{b'})}\nonumber\\
&\times&e^{-2\eta(\beta_b^2+\dot{\beta}_b^2)(1-z_bz_{b'})}e^{-2\eta(\beta_a^2+\dot{\beta}_a^2)(1-z_az_{a'})}\nonumber\\
&\times&\mathcal{D}(0),
\label{totalef}
\end{eqnarray}
where the first term is the unitary conditional two-qubit phase and the next two are single qubit dephasing terms. The corresponding operation undergone by the qubits, omitting the conditional phase is 

\begin{eqnarray}
\tilde{\mathcal{S}}\mathcal{S}\rho &=&(1-p_a)(1-p_b)\rho+p_a(1-p_b)Z_a\rho Z_a\nonumber\\
&+&(1-p_a)p_bZ_b\rho Z_b+p_ap_bZ_aZ_b\rho Z_bZ_a,
\label{ezmap}
\end{eqnarray}
with

\begin{equation}
p_{a|b}=\frac{1-e^{-4\eta(\beta_{a|b}^2+\dot{\beta}_{a|b}^2)}}{2},
\label{p}
\end{equation}
the probability that each qubit incured a $Z$ error. Clearly these dephasing processes are independent, leading to a very simple operation, in contrast with the partial operations (31) or even (32) obtained in the single interaction sequence. Due to the fact that the geometrical phase and the qubit dephasing both double for a single iteration of the gate, there is no advantage to further reducing the sizes of the displacements and increasing the number of iterations. 

The output state fidelity and entanglement of this sequence are plotted in Fig. 7(b), setting $\beta_a=\beta_b=\sqrt{\pi/8(e^{-l}+e^{-3l})}$. The fidelity is very similar while the concurrence is slightly reduced, compared to the single sequence. So for entanglement distribution in view of communication applications, which only follow the purification of an entangled state up to an acceptable level, the iterated scheme is penalizing, as it requires twice the amount of time and reduces the entanglement. But in view of full blown quantum computation with quantum error correction, the iterated scheme may present a serious advantage, suppressing correlated errors at the gate level.

\section{V. discussion}
In this work we set out to examine the vital issue of dissipation in a CV quantum bus being used to realize quantum information processing tasks over discrete subsystems. We first looked at dissipation in the probe during interactions. We considered a general interaction Hamiltonian between continuous and discrete subsystems which appeared in two different physical settings: all optical cross-phase and dispersive CQED settings. Solving the master equation we derived the closed form of the coherence parameter (7) discussed in some previous work \cite{jacob}. This parameter can be expressed as a function of $\alpha$, $\gamma/\chi$ and $\chi t$ and has a non-zero limiting behavior for large interaction times. As expected, it has the effect of a phase flip channel (10) in the case that the subsystem is a qubit. More interestingly, loss on the probe mode induces a conditional phase operation on the qubit, which needs to be undone in the case that the interaction is part of a more elaborate gate. The resulting time dependent density matrix we were able to rewrite as a two-qubit density matrix. From this we observed the evolution of the entropy and the entanglement between the continuous variable and the qubit during the interaction. Increasing the probe amplitude results in higher entanglement with a sharper peaking, meaning the interaction time has to be controlled more and more precisely. We also found that increasing the probe amplitude increases the coherence of the combined state at entanglement peaking time. 

This could be a possible approach to minimizing decoherence effects in qubus gates, however it has to be balanced against loss in between interactions and the conditional phase discussed above which depends strongly on the amplitude. As we saw in sections III and IV the dephasing effect on the qubit is related to the overlap between the possible states of the probe in phase space, which is a function of the squared distance between them. Thus increasing $\alpha$ in the interactions will augment the effects of loss in the transmission. The solution to this trade-off will depend on the physical system at hand. 

Having solved the effects of loss on interactions, we combined them with displacements to simulate conditional displacements. These couplings are fundamental to hybrid schemes making use of the geometrical phase acquired by the probe in phase space. We managed to compensate for the combined dephasing effects during and in between interactions, obtaining the desired coupling up to a single qubit dephasing. This result is crucial to the analysis of the full CZ gate undertaken in section IV, in which we compounded the conditional displacement operations and added loss in between them too. The first and last loss terms constitute single qubit errors whereas the middle term contains correlated errors. We calculated the resulting quantum operation on the pair of qubits and found that in the low loss regime it only contained single qubit dephasing errors, given a QEC setting with syndrome measurements. To gain more insight into the gate's performance, we looked at the output state fidelity, for equally weighted input superposition states. We found that the fidelity remained high and that even for losses of up to 8 dB in probe signal, the output states could be purified. A simple iteration procedure then allowed us to remove extra correlated errors altogether, presenting a serious advantage in view of a full-blown quantum computing scheme. This full-blown scheme will also have to factor other sources of errors such direct qubit decoherence processes.

The method we use here is very broad and could be applied to general bus mediated quantum information processing schemes. Loss of coherence in the bus will automatically result in dephasing on the qubits coupled to it at that time. If several qubits are simultaneously coupled then correlated errors will arise. These will have to be minimized in order to efficiently correct for errors. However the peculiarity of the qubus scheme is that we are using non-orthogonal states of the bus to encode information held by the subsystems. Surprisingly, in the absence of loss, perfect multiqubit gates can be implemented this way. In a dissipative setting, this non-orthogonality leads to overlap calculations which are very likely to arise in general non-orthogonal bus schemes.

Following the results presented in this paper, there are two main directions for future work. The first is to find a way of dealing with the single qubit phases accumulated throughout qubus gates. Simply undoing them would require considerable precision and this may not be a realistic option. The second is to investigate possible schemes to reduce dephasing effects, possibly by engineering the probe itself.

\section{Acknowledgements} We would like to thank S. Devitt, T.D. Ladd, P. van Loock, N. L\"{u}tkenhaus and T. Karasawa for useful insights and discussions. SGRL acknowledges the support of a Monbukagakusho scholarship. This work was supported in part by MEXT in Japan and the EU project QAP.

\begin{center}
    {\bf APPENDIX}
  \end{center}

The exponent $f_{nm}$ in the coherence parameter $\zeta_{nm}=e^{f_{nm}}$ can be written as $\mathrm{Re}[f_{nm}]+i\mathrm{Im}[f_{nm}]$. Here we give expressions for the real and imaginary parts of the exponent with $\hat{\Lambda}=Z$. Beginning with the real part we obtain

\begin{eqnarray}
e^{\mathrm{Re}[f_{nm}]}&=&\mathrm{exp}[-\frac{|\alpha|^2}{2(\gamma^2+\chi^2)}(\chi^2(1-e^{-2\gamma t})\nonumber\\
&-&2\gamma^2e^{-2\gamma t}\mathrm{sin}^2\chi t-\chi\gamma e^{-2\gamma t}\mathrm{sin}2\chi t)\nonumber\\
&\times &(1-z_nz_m)],
\label{}
\end{eqnarray}
representing the decay of the off-diagonal components of the density matrix. In the limit of large interaction times this term leads to a fixed dephasing effect
\begin{equation}
e^{\mathrm{Re}[f_{nm}]}_{t\rightarrow\infty}=\mathrm{exp}\left[-\frac{|\alpha|^2}{2(1+(\gamma/\chi)^2)}(1-z_nz_m)\right].
\label{}
\end{equation}

Then the imaginary part of the exponent is
\begin{eqnarray}
e^{\mathrm{Im}[f]}&=&\mathrm{exp}[\frac{i\gamma|\alpha|^2}{2(\gamma^2+\chi^2)}(\chi(1-e^{-2\gamma t}\mathrm{cos}2\chi t)\nonumber\\
&-&\gamma\mathrm{sin}2\chi t)(z_n-z_m)],
\label{}
\end{eqnarray}
corresponding to a known single qubit phase term.


\begin{thebibliography}{99}
\bibitem{ekert} A. Ekert and R. Jozsa, Rev. Mod. Phys. \textbf{68}, 733 (1996).
\bibitem{lloyd1} S. Lloyd, Science \textbf{273}, 1073 (1996).
\bibitem{abrams} D.S. Abrams and S. Lloyd, Phys. Rev. Lett. \textbf{79}, 2586 (1997).
\bibitem{cirac} J.I. Cirac and P. Zoller, Nature (London) \textbf{404}, 579 (2000).
\bibitem{klm} E. Knill, R. Laflamme and G. Milburn, Nature (London) \textbf{409}, 46 (2001).
\bibitem{kane} B.E. Kane, Nature (London) \textbf{393}, 133 (1998).
\bibitem{laddsil} T.D. Ladd, J.R. Goldman, F. Yamaguchi, Y. Yamamoto, E. Abe and K.M. Itoh, Phys. Rev. Lett. \textbf{89}, 017901 (2002).
\bibitem{briegel} H.J. Briegel and R. Raussendorf, Phys. Rev. Lett. \textbf{86}, 910 (2001).
\bibitem{vander} L.M.K. Vandersypen et al., Nature \textbf{414}, 883 (2001).
\bibitem{sack} C.A. Sackett et al., Nature \textbf{404}, 256 (2000).
\bibitem{pit} T.B. Pittman, M.J. Fitch, B.C Jacobs, J.D. Franson, Rev. A \textbf{68}, 032316 (2003).
\bibitem{walther} P.Walther et al., Nature \textbf{434}, 169 (2005).
\bibitem{div} D.P. DiVincenzo, Fortschr. Phys. \textbf{48}, 771 (2000).
\bibitem{plen-cqed} M.B. Plenio, S.F. Huelga, A. Beige, and P.L. Knight, Phys. Rev. A \textbf{59}, 2468 (1999).
\bibitem{andy} A.D. Greentree, J.H. Cole, A.R. Hamilton, and L.C.L. Hollenberg, Phys. Rev. B \textbf{70}, 235317 (2004).
\bibitem{tim} T.P. Spiller, K. Nemoto, S.L. Braunstein, W.J. Munro, P. van Loock and G.J. Milburn, New J. Phys. \textbf{8}, 30 (2006).
\bibitem{sam} S. Lloyd and S.L. Braunstein, Phys. Rev. Lett. \textbf{82}, 1784 (1999).
\bibitem{loockrev} S.L. Braunstein and P. van Loock, Rev. Mod. Phys. \textbf{77}, 513 (2005).
\bibitem{lloyd} S. Lloyd, quant-ph/0008057 (1999).
\bibitem{wang} X. Wang, A. S\o rensen and K. M\o lmer, Phys. Rev. Lett. \textbf{86}, 3907 (2001).
\bibitem{zan} X. Wang and P. Zanardi, Phys. Rev. A \textbf{65}, 032327 (2002).
\bibitem{mil} G.J. Milburn, quant-ph/9908037 (1999).
\bibitem{qndmil} G.J. Milburn and D.F. Walls, Phys. Rev. A \textbf{30}, 56 (1984).
\bibitem{qndimoto} N. Imoto, H.A. Haus, and Y. Yamamoto, Phys. Rev. A \textbf{32}, 2287 (1985).
\bibitem{mun} W.J. Munro, K. Nemoto, T.P. Spiller, S.D. Barrett, P. Kok, and R.G. Beausoleil, J. Opt. B: Quantum Semiclass. Opt. \textbf{7} S135 (2005).
\bibitem{mun2} W.J. Munro, K. Nemoto and T.P. Spiller, New J. Phys. \textbf{7}, 137 (2005).
\bibitem{seb1} S.G.R. Louis, K. Nemoto, W.J. Munro and T.P. Spiller, Phys. Rev. A \textbf{75} 042323 (2006). 
\bibitem{kae} K. Nemoto and W.J. Munro, Phys. Rev. Lett. \textbf{93}, 250502 (2004).
\bibitem{seb2} S.G.R. Louis, K. Nemoto, W.J. Munro and T.P. Spiller, New J. Phys. \textbf{9}, 193 (2007).
\bibitem{loock-rep} P. van Loock, T.D. Ladd, K. Sanaka, F. Yamaguchi, K. Nemoto, W.J. Munro, and Y. Yamamoto, Phys. Rev. Lett. \textbf{96}, 240501 (2006).
\bibitem{ladd} T.D. Ladd, P. van Loock, K. Nemoto, W.J. Munro, Y. Yamamoto, New J. Phys. \textbf{8}, 164 (2006).
\bibitem{jay} E.T. Jaynes and F.W. Cummings, Proc. IEEE \textbf{51}, 89 (1963).
\bibitem{tim2} D.A. Rodrigues, C.E. A. Jarvis, B.L. Gy\"{o}rffy, T.P. Spiller, and J.F. Annett, J. Phys.: Condens. Matter \textbf{20}, 075211 (2008).
\bibitem{wallraff1}
A. Wallraff, D.I. Schuster, A. Blais, L. Frunzio, R.-S. Huang, J. Majer, S. Kumar, S.M. Girvin, and R.J. Schoelkopf, Nature \textbf{431}, 162 (2004). 
\bibitem{wallraff2}
A. Wallraff, D.I. Schuster, A. Blais, L. Frunzio, J. Majer, M.H. Devoret, S.M. Girvin, and R.J. Schoelkopf, Phys. Rev. Lett. \textbf{95}, 060501 (2005).
\bibitem{nemes} J.G.P. de Faria and M.C. Nemes, Phys. Rev. A \textbf{59}, 3918 (1999).
\bibitem{rom} K.M.F. Romero and M. C. Nemes, Physica A \textbf{325}, 333 (2003).
\bibitem{bar} S.D. Barrett and G.J. Milburn, Phys. Rev. A \textbf{74}, 060302(R) (2006).
\bibitem{cen} L.-X. Cen and P. Zanardi, Phys. Rev. A \textbf{71}, 060307(R) (2005).
\bibitem{jacob} H. Jeong, Phys. Rev. A \textbf{72}, 034305 (2005).
\bibitem{jacob2} H. Jeong, Phys. Rev. A \textbf{73}, 052320 (2006).
\bibitem{knight} C.C. Gerry and P.L. Knight, \textit{Introductory Quantum Optics}, Cambride University press (2005).
\bibitem{gardiner} C.W. Gardiner and P. Zoller, \textit{Quantum noise}, Springer (2000).
\bibitem{wit} W. Witschel, Int. J. Quantum Chem. \textbf{20}, 1233 (1981).
\bibitem{niel} M.A. Nielsen and I.L. Chuang, \textit{Quantum Computation and Quantum Information}, Cambride University press (2000). 
\bibitem{norb} N. L\"{u}tkenhaus and P. van Loock, private communication.
\bibitem{plen-ent-rev} M.B. Plenio and S. Virmani, Quant. Inf. Comp. \textbf{7}, 1 (2007).
\bibitem{woot} C.H. Bennet, D. P. DiVincenzo, J. Smolin and W. K. Wooters, Phys. Rev. A \textbf{54}, 3824 (1996).
\bibitem{woot2} W.K. Wooters, Phys. Rev. Lett. \textbf{80} 2245 (1997).
\bibitem{mil2} G.J. Milburn and D.F. Walls, Phys. Rev. A \textbf{30}, 56 (1984).
\bibitem{imo} N. Imoto, H.A. Haus, and Y. Yamamoto, Phys. Rev. A \textbf{32}, 2287 (1985).
\bibitem{blais} A. Blais, R. Huang, A. Wallraff, S.M. Girvin and R.J. Schoelkopf, Phys. Rev. A \textbf{69}, 062320 (2004).
\bibitem{billprep} W.J. Munro, in preparation.
\bibitem{loock} P. van Loock, W.J. Munro, K. Nemoto, T.P. Spiller, T.D. Ladd, S.L. Braunstein and G.J. Milburn, quant-ph/0701057 (2007).
\bibitem{walls} D.F. Walls and G.J. Milburn, Phys. Rev. A \textbf{31}, 2403 (1985). 
\bibitem{dev} S. Devitt, private communication.





\end{thebibliography}
\end{document}